\documentstyle[a4,aps,prb,graphics]{revtex}
\begin{document}
\large

\title{\Large {\em Ab initio} evaluation of local effective interactions \\
in $\alpha^\prime N\!aV_2O_5$}

\author{Nicolas Suaud and Marie-Bernadette Lepetit}

\address
{Laboratoire de Chimie Quantique et Physique Mol\'eculaire, \\
Unit\'e Mixte de Recherche 5626 du CNRS,\\
118 route de  Narbonne, F-31062 Toulouse, France}

\date{\today}

\maketitle
\vspace*{1cm} 
\begin{abstract}
We will present the numerical evaluation of the hopping and magnetic
exchange integrals for a nearest-neighbor $t-J$ model of the
quarter-filled $\alpha^\prime N\!aV_2O_5$ compound.  The effective
integrals are obtained from valence-spectroscopy {\em ab initio}
calculations of embedded crystal fragments (two $VO_5$ pyramids in the
different geometries corresponding to the desired parameters). We are
using a large configurations interaction (CI) method, where the CI
space is specifically optimized to obtain accurate energy differences.
We show that the $\alpha^\prime N\!aV_2O_5$ system can be seen as a
two-dimensional asymmetric triangular Heisenberg lattice where the
effective sites represent delocalized $V-O-V$ rung entities supporting
the magnetic electrons.
\end{abstract}


\large
\section{INTRODUCTION}
Spin-Peierls (SP) transitions have been long known in organic
compounds~\cite{Bray} but the observation, in 1993, of a SP transition
in the inorganic $CuGeO_3$~\cite{Hase} ($T_{SP}=14K$) renewed the
interest of the physical community for this phenomenum. In 1996 the
$\alpha^\prime N\!aV_2O_5$~\cite{Hardy} inorganic system was found to
present a second order SP transition~\cite{Isobe} at
$T_{SP}=34K$, the highest $T\small_{SP}$ temperature so far
known. Indeed a rapid drop of the magnetic susceptibility, the opening
of a spin gap~\cite{gap} ($\Delta=9.8meV$) as well as the
magneto-distortion of the lattice was observed in this compound. This
compound has attracted special attention since, in addition of its
very high $T_{SP}$, the value of the
$\frac{2\Delta}{k\small_BT\small_{SP}}=6.44$ ratio does not agree with
the BCS theory predicted value of $3.53$, the $T_{SP}$
dependence with the magnetic field is much weaker~\cite{Kremer} than
the theoretical predictions~\cite{Bulaevskii} and the temperature
dependence of the thermal conductivity is much larger than in
$CuGeO_3$~\cite{Vasilev}.

\begin{figure}[h]
\hspace*{-0.50cm}
\includegraphics{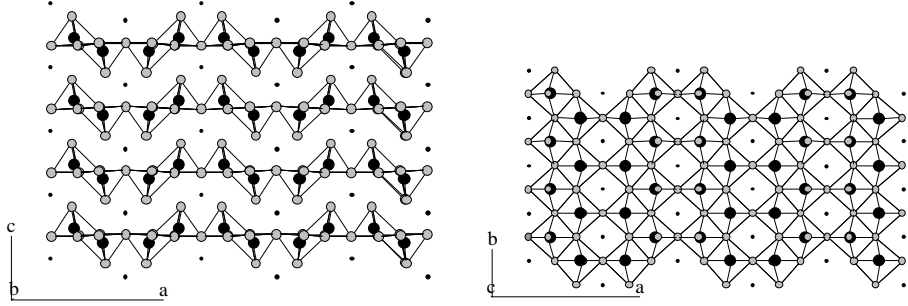} 
\caption{ Schematic structure of $\alpha^{'}N\!aV_2O_5$, a) along the
a-c plane, b) along the a-b plane. The oxygen atoms are are denoted by
open circles, the vanadium atoms by filled circles and the sodium
atoms by dots.}
\label{Fig1}
\end{figure}

The crystal is formed by layers of $VO_5$ square-pyramids stacked
along the $c$ axis. The oxygen atoms of the pyramid basis form a
quasi-square planar lattice along the $a$ and $b$ directions.  The
pyramids are alternatively pointing on top or below these
planes. Periodical vacancies of the $VO$ top of the pyramids are
replaced by $Na^+$ ions forming chains parallel to the b axis (see
fig.~\ref{Fig1}). 


For a long time, the high temperature crystal structure was assumed to
be non-centrosymmetric (in $P2_1mn$ group)~\cite{Galy}, and
consequently the electronic structure was assumed to be an alternation
of half filled $V^{4+}$ magnetic chains and of non-magnetic $V^{5+}$
chains along the b axis. The system was then supposed to be
constituted of isolated Heisenberg chains and the SP
transition was naturally understood in this description by the 
one-dimensional dimerization of the $V^{4+}$ magnetic chains.

However the existence of a pseudo-inversion center in this compound
led a couple of groups~\cite{Schnering} to re-investigate the
crystallographical data using more recent experimental methods. They
found  that $\alpha^\prime N\!aV_2O_5$ was centrosymmetric (in
the $Pmmn$ space group) destroying the grounding for the assumed
differentiation between $V^{4+}$ and $V^{5+}$ chains. $\alpha^\prime
N\!aV_2O_5$ should then be seen as a quarter-filled system. 

It is widely accepted, after Galy~\cite{Galy2}, that the magnetic
interactions take only place within the layers, and thus the pertinent
model for $\alpha^\prime N\!aV_2O_5$ is a quarter-filled $t-J$ model
Hamiltonian in the $(a,b)$ plane, based on the $V$ magnetic $3d$
orbitals, where the interactions are limited to the nearest neighbors (NN)
magnetic sites. 
\begin{eqnarray} 
H&=& \sum_{<i,j>} t_{ij} \sum_{\sigma}\left(
 c^\dagger_{i_\sigma}c_{j_\sigma} +
 c^\dagger_{j_\sigma}c_{i_\sigma}\right) - \sum_{<i,j>}
 J_{ij}\left(\vec{S_i}.\vec{S_j} - {1 \over 4} n_i n_j\right)
\end{eqnarray} 
where the sum over $<i,j>$ runs over NN magnetic sites, $\vec{S_i}$ is
the local spin operator on site $i$, $c^\dagger_{i_\sigma}$
(resp. $c_{i_\sigma}$) are the usual creation (resp. annihilation)
operators of an electron of spin $\sigma$ on site i, $n_i$ is the
number operator on site $i$, $J_{ij}$ is the magnetic exchange
integral and $t_{ij}$ the hopping integral of a magnetic electron
between sites i and j.

\begin{figure}[h]
\hspace{3cm}
\includegraphics{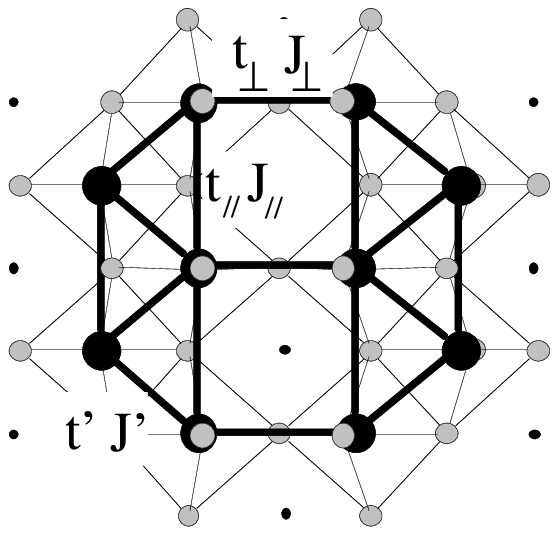} 
\caption{The three different types of NN interactions: $t_\|$ and  $J_\|$,
$t_\bot$ and $J_\|$, $t^\prime$ and $J^\prime$.}
\label{Fig2}
\end{figure}

In the present system one has three different types of NN interactions
(see fig.~\ref{Fig2})~: the interactions along the ladders (in the $b$
direction) denoted as $t_{\|}$ and $J_{\|}$, the interactions along
ladders rungs denoted as $t_{\bot}$ and $J_{\bot}$ (in the $a$
direction) and the interactions between the ladders denoted as
$t^{\prime}$ and $J^{\prime}$.  According to the order of magnitude of
these parameters, the system electronic structure can be very
different, going from a one dimensional behavior with essentially non
interacting two-legs ladders ($|t^\prime|, |J^\prime| \ll 1$), or
zig-zag frustrated ladders ($|t_\bot|, |J_\bot| \ll 1$) to a
two-dimensional system. It is therefore of crucial importance to have
a reliable evaluation of the relative amplitude of the different parameters.

The aim of this work is to evaluate, using {\em ab initio} quantum
chemical methods, the effective hopping and exchange integrals between
NN vanadium atoms.

\section{METHOD}

The exchange and hopping effective integrals are essentially local
parameters, therefore they can be accurately evaluated using fragment
spectroscopy, provided that (i) the fragment includes all the crystal
short range effects, that is the local environment of the magnetic
atoms, and (ii) the crystal long range effects are treated through an
appropriate bath.

\begin{center}
\begin{figure}[h]
\resizebox{6cm}{6cm}{\includegraphics{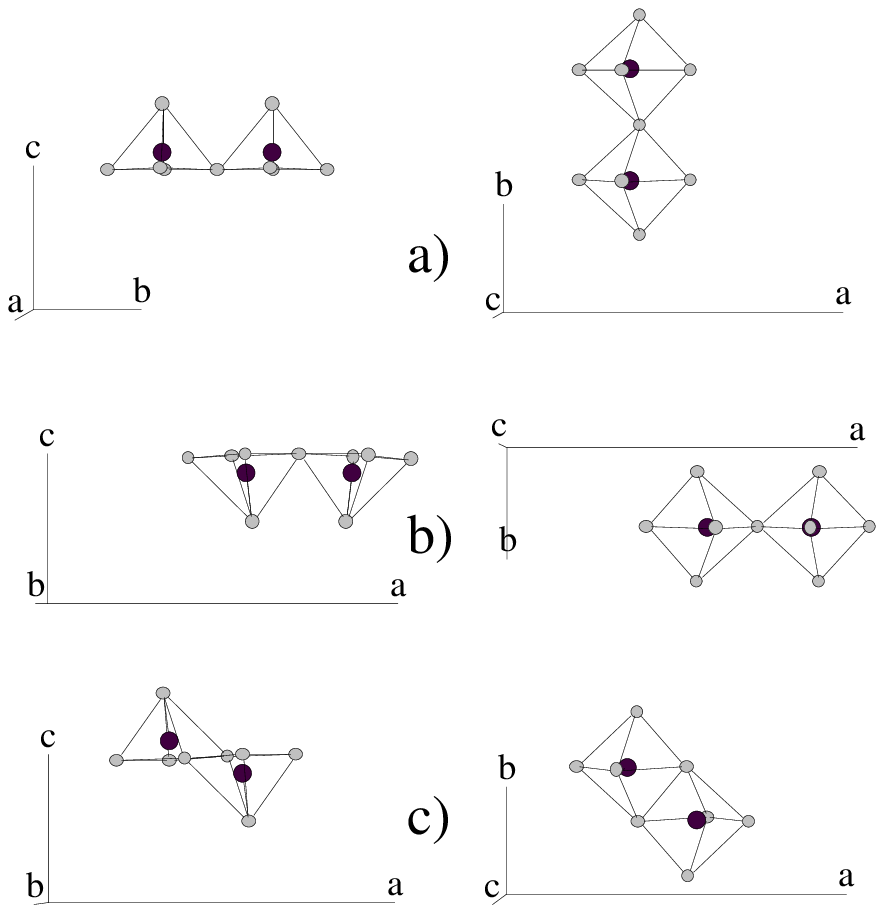} }
\caption{Fragments of the crystal used for the cluster
calculations. a) $t_\|, \, {\rm and}\, J_\|$~; b) $t_\bot,\, {\rm
and}\, J_\bot$~; c) $t^\prime,\, {\rm and} \, J^\prime$.}
\label{Fig3}
\end{figure}
\end{center}

We performed excitation energies calculations on
bi-pyramidal fragments (including the two NN magnetic atoms as well as
all their surrounding oxygens) in a bath reproducing the main effects of
the rest of the crystal (Madelung potential and exclusion effects).
The fragments used for the $t_\|$ (resp.  $J_\|$) and $t_\bot$ (resp.
$J_\bot$) calculations are built from two $VO_5$ pyramids sharing a
corner (see fig.~\ref{Fig3} a and b) while the fragment used for the
$t^\prime$ (resp.  $J^\prime$) calculations is constituted of two
$VO_5$ pyramids sharing an edge (see fig.~\ref{Fig3} c).  
Two sets of
calculations are performed for each fragment. The first one involving
two unpaired electrons in the two $VO_5$ pyramids allows us to obtain
the effective exchange parameters (including all direct and
super-exchange processes) from the singlet-triplet first excitation energy. 
\begin{eqnarray}
J&=& E_S-E_T
\end{eqnarray} 
The second set of calculations, involving only one unpaired magnetic
electron in the two $VO_5$ pyramids, allows us to obtain the
effective hopping parameters from the first doublet-doublet excitation
energy.
\begin{eqnarray}
t&=& \frac{E_{D_{+}}-E_{D_{-}}}{2}
\end{eqnarray} 
where the $D_{+}$ (resp. $D_{-}$) doublet is of the same symmetry than 
the $3d_1 + 3d_2$ (resp. $3d_1 - 3d_2$) delocalized orbital on the two
magnetic centers. One notes that according to the relative energies of
the $D_{+}$ and $D_{-}$ states, the sign of the hopping integral can
change.

\subsection{Modeling the rest of the crystal}

As state above a correct fragment calculation should take into account
the main effects of the rest of the crystal. In our ionic, strongly
correlated system these effects are limited to the short range Pauli
exclusion effects and the long range Madelung potential.

First, the Madelung potential is reproduced by a set of positive and
negative charges corresponding to the cations and anions. We used a
cutoff threshold associated with a $13\AA$~radius sphere, centered on
the bi-pyramid fragment. The border of the sphere is designed so that
to preserve $VO_5$ pyramids, the chemical meaningful entities.  Border
charges are adapted by an Evjen procedure. 

The $\alpha^\prime N\!aV_2O_5$ system presents $4$ different types of
atoms~: the fully ionized sodium cations, the mixed valence vanadium
atoms and two types of oxygen atoms, the in-plane oxygens and the
apical oxygens. These two types of oxygen play a very different role
in the chemistry of the compound. Indeed, ab-initio calculations on
the bi-pyramidal systems, both at the Hartree-Fock (HF) and correlated
level, as well HF calculations on the whole crystal give the same
result, the apical oxygen is bonded to the vanadium atom by a slightly
polarized double bond. The M\"ulliken population analysis are similar
in all calculations with a charge close to $-2\bar{e}$ for the
in-plane oxygens while the apical ones present only a small charge of
about $-0.5\bar{e}$. In fact the $VO_5$ pyramids should not be seen as
such, but rather as $(V=O)^{2+\;{\rm or}\;3+}$ molecules on an oxygen
double-anions square lattice. This picture is confirmed by Raman
experiments~\cite{bacsa}.  Indeed, two distinct bands are observed,
one for the stretch of the vanadium/apical-oxygen bond located at
$970\, {\rm cm}^{-1}$, the other for the group of vibrations of the
vanadium against the oxygen plane which is located in the $400$-$480\,
{\rm cm}^{-1}$ range.

The embedding charges are then taken as follow~:
\begin{itemize} 
\item the sodium atoms are represented by $+1.0\bar{e}$ charges, 
\item the oxygen atoms of the quasi-square lattice forming the basis of the pyramids are represented by $-2.0\bar{e}$ charges, 
\item the apical oxygens are represented by $-0.5\bar{e}$ charges 
\item and the vanadium atoms are represented by their average valence
of $+3.0\bar{e}$. 
\end{itemize}

Second, the exclusion effects are taken into account through total
charges pseudo-potentials. The embedded fragments are highly
negatively charged ($-8\bar{e}$ or $-9\bar{e}$) and their electrons
would like to expand out of the fragments volume, thus the contention
effect of the rest of the crystal needs to be modeled by an exclusion
potential. The negative charge of the oxygen anions, even modeled as
point charges, is sufficient to insure the exclusion effect on the
fragment electrons~; however the positive charge of the cations
attracts the fragment electrons and it is necessary to explicitly
treat the exclusion. This is done by modeling the cations with
total-ions pseudo-potentials~\cite{total}, using the Durand and
Barthelat large core effective potentials~\cite{Durand}, instead of
only point charges.

\begin{figure}[h]
 \begin{minipage}{5cm}
   \centerline{\resizebox{5cm}{8cm}{\includegraphics{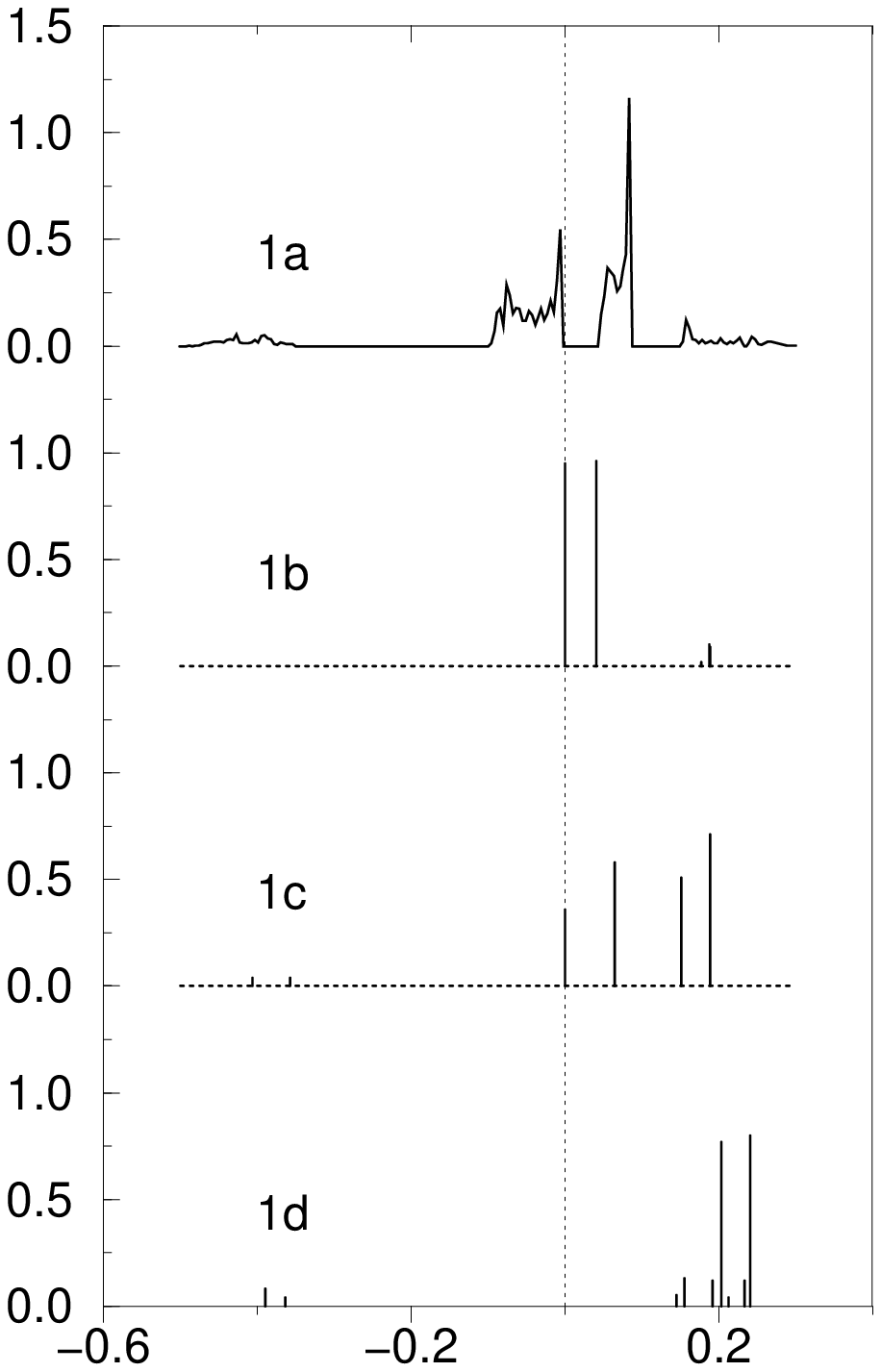}}}
\end{minipage} \hspace*{4eM}
   \begin{minipage}{5cm}
   \centerline{\resizebox{5cm}{8cm}{\includegraphics{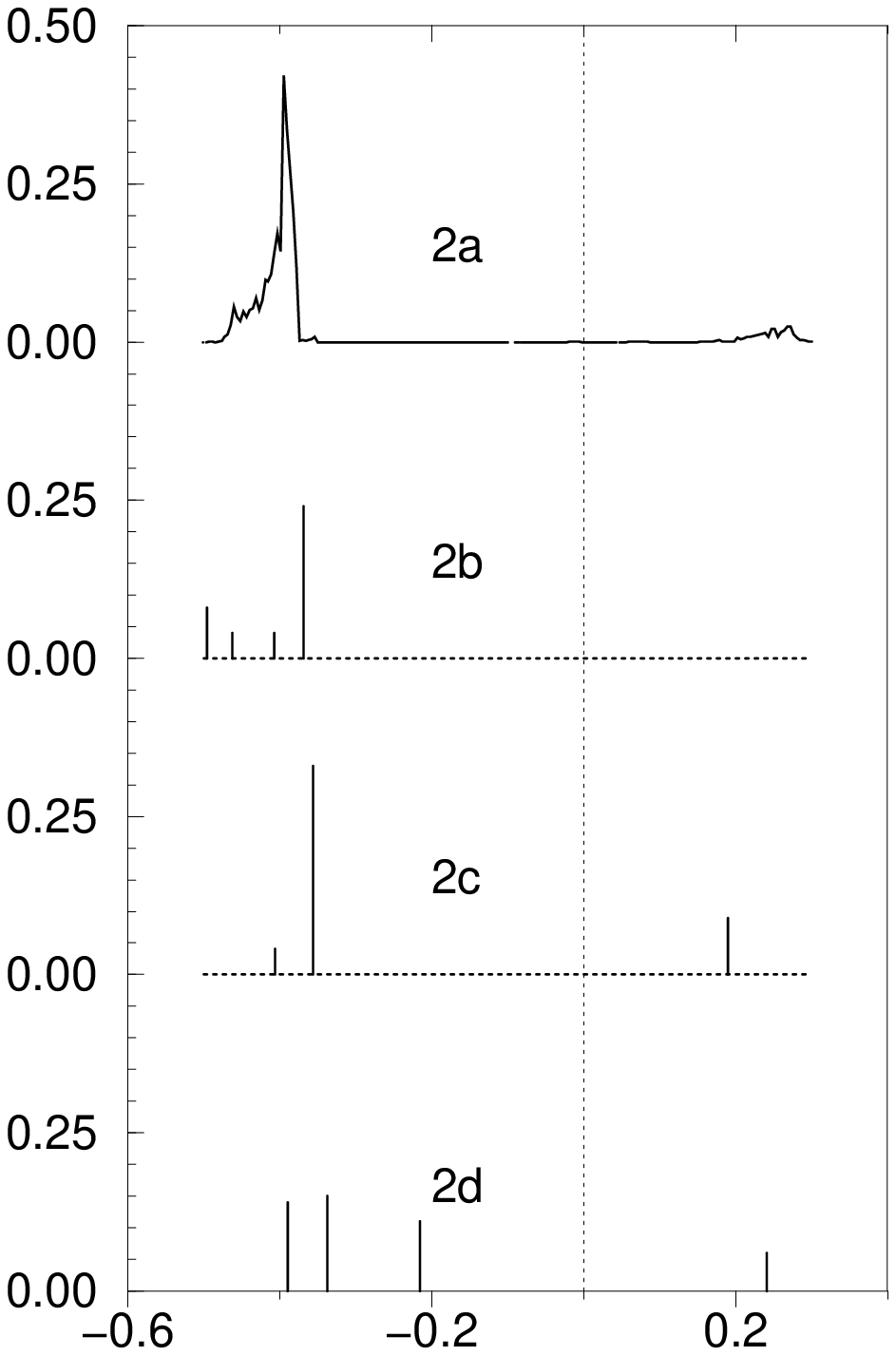}}}
 \end{minipage} \vspace{1eM}
\caption{Projected DOSS on the $d$ magnetic atomic orbitals (1) and
the $p$ orbital of the bridging oxygen atom in the rung geometry (2).
The crystal HF calculation (a). The rung cluster of this work (b). The
rung cluster embedded with only one type of oxygen atom, all atoms
supporting their formal charges (c). The rung cluster without
embedding (d). The dotted vertical line shows the Fermi level.}
\label{Fig:doss} 
\end{figure}

The good behavior of the above embedding has been checked against HF
crystal calculations. Figure~\ref{Fig:doss} shows the
Projected Density of States (PDOSS) on the magnetic orbitals of the
vanadium atoms (1) and on the $p$ orbital of the bridging oxygen of
the rungs (2). The PDOSS is compared with its equivalent in the
cluster calculations, the square of the cluster orbitals projection
onto the atomic orbitals, plotted as a function of the cluster orbital
energies. On can see that while the above embedding yields a favorable
comparison of the cluster orbitals with the crystal PDOSS, the
embedding defined with only one kind of oxygen atoms and the formal
charges ($O^{-2}$, $V^{+4.5}$, $N\!a^+$) do not properly place the
vanadium $d$ orbitals, in particular the gap at the Fermi level is
overestimated. In order to emphasize the importance of a proper
embedding, the isolated cluster orbitals are also reported. The result
goes without comment.


\subsection{Computing accurate excitation energies}

{\it Ab initio} quantum chemical methods are powerful
tools to obtain reliable excitation energies as well as good
representation of the associated ground or excited states wave
functions.  As will be seen later a precise analysis of the wave
function informations allows a deep understanding of the different
mechanisms involved in the effective exchange and hopping processes.

Let us first consider the physics of the excitations we are trying to
compute. One sees easily that the fragment orbitals can be divided
into three subsets according to their role in the many-body wave
functions of the two states we are looking for.
\begin{description}
\item[The occupied orbitals] are the orbitals that always remain 
doubly-occupied in the many-body states. 
\item[The active orbitals] are the ones which occupation number can
change in the different Slater determinants participating to the
many-body states. In our case these orbitals are not only the two $3d$
magnetic orbitals (from now on we will refer to them as $d_1$ and
$d_2$ according to the $V$ atom to which they belong) located on the
two vanadium atoms but also the $2p$ orbitals (we will refer to them
as $p$) of the bridging oxygen atom(s) that can participate to the
effective exchange through super-exchange mechanism and to the hopping
process by a through bridge delocalization of the hole. Symmetry
considerations reduce the number of these bridging orbitals to two $2p$
orbitals (one per irrep. involved). One should note that these active
orbitals correspond to the valence orbitals that would support a
three-bands Hubbard representation of the system ---~where both the $V$ and
the $O$ atoms of the pyramids bases are considered.
\item[The virtual orbitals] are the orbitals that always remain 
empty in the many-body states. 
\end{description}

A good ``zeroth-order'' description of the seeked states will then be
provided by the eigensolutions of the exact Hamiltonian in the
Complete Active Space (CAS)~; the CAS is defined as all possible
configurations built from the occupation rules given above. In fact we
simultaneously optimize the orbitals and the wave function
coefficients in a Complete Active Space Self Consistent Field (CASSCF)
procedure~\cite{CASSCF}.  At this level of calculation, the
polarization and correlation effects within the active space (static
polarization and correlation) are variationally taken into account,
while the other electrons are described within a mean field
approximation.  In the $J$ calculations, six electrons ---~the two
unpaired electrons plus the four electrons originating from the
bridging $p$ orbitals of the central oxygen(s)~--- are distributed in
all possible manners into the four active orbitals. In $t$
calculations, only five electrons are distributed into the four active
orbitals.

``Dynamical'' polarization and correlation effects coming from
excitations out of the active shell (originating from the occupied
orbitals or ending in the virtual ones) are however crucial in order
to obtain reliable results for excitation energies.  The ``dynamical''
polarization effects come from the single-excitations on the CASSCF
wave function, while the dominant contributions to ``dynamical''
correlation effects come from the double-excitations on the CASSCF
wave function. We will take these effects into account in a two steps
procedure. \\ a) A self-consistent procedure is performed. It involves
the diagonalization of the CAS + all single excitations space as well
as a new orbitals optimization at this level. This is the so-called
Iterative Difference Dedicated Configuration Interaction level 1
method (IDDCI1)~\cite{IDDCI}. \\ b) In a second step the double
excitations on the CAS wave-function that participate to the
energy difference (at the $2^{nd}$ order of perturbation) are added to
the CI space, which is diagonalized. This is the Difference Dedicated
Configuration Interaction level 2 (DDCI2)~\cite{DDCI}.

The above method is powerful and reliable. It has been used with great
success in a slightly different version (DDCI3) for the determination
of the $t-J$ model parameters, for instance, on
copper-oxides~\cite{cuo} or ladder compounds such as $S\!r C\!u_2
O_3$~\cite{sr}. The main differences between the procedure used in the
present work and the works cited above are i) the optimization of the
orbitals at the CAS + single excitations level (IDDCI1) instead of
only at the CASSCF level ii) the inclusion of the dynamical
polarization and correlation of the bridging oxygen $p$ orbitals is
done in a slightly different way. We choose to include the active orbital(s)
of the bridge in the CAS so that the dynamical polarization effects
are taken into account in the DDCI1 space and the dynamical
correlation effects are taken into account in the DDCI2 space. De
Graaf {\it et al}~\cite{sr} as well as Calzado {et al}~\cite{cuo} made
the choice of a smaller CAS including only the two magnetic $d$
orbitals and to extend the diagonalization to the 2-holes 1-particle
and 2-particles 1-hole CI space (DDCI3) in order to describe the same
physics. The main advantages of our choice are the following \begin{itemize}
\item The optimization of the orbitals at a level including the dynamical polarization effects.
\item An equal treatment of the magnetic $d$ orbitals and the active
$p$ orbitals of the bridge. This point can be important in the cases
where the through bridge processes are large.
\item A much smaller size of the largest CI space to diagonalize. For
instance in the calculation of the $J_\bot$ effective integral our
choice yields a CI space of $592,000$ determinants while the DCCI3
space with the small CAS would give a CI of $2,783,000$ determinants.
\end{itemize}

In order to analyze the relative contributions of dynamical
polarization and correlation we performed an additional Configuration
Interaction, namely the diagonalization of the CAS but defined on the
set of orbitals optimized within the IDDCI1 procedure (CASCI).

\subsection{Computational details}
All cluster calculations are performed using the Barandiaran
basis sets~\cite{Barandiaran} with the recommended contraction:
$3s3p4d$ for $V$ atoms and $2s4p1d$ for $O$ atoms. The effects of the
inner electrons are modeled by core pseudo-potentials~\cite{Barandiaran},
$[1s^22s^22p^63s^2]$ for $V$ atoms and $[1s^2]$ for $O$ atoms.

The crystal HF calculations were performed using the CRYSTAL98
package~\cite{crystal}. In order to be able to perform the infinite
crystal calculation it has been necessary to reduce the basis set
compared to the cluster calculations. We therefore used a basis
set~\cite{base} of quality single-zeta for the core electrons,
while the valence electrons were described using a double-zeta, plus
one polarization function for the oxygen and two polarization
functions for the vanadium.

\section{RESULTS AND DISCUSSION}

The computed hopping and exchange integrals are summarized in
table~\ref{tab:tJres}. 

The differences observed between the CASCI and the IDDCI1 results
point out the crucial importance of the inclusion of dynamical
{\em polarization} effects in order to obtain reliable results for effective
exchange and hopping processes. In comparison the dynamical
correlation effects are of much lesser importance since they modify
the integrals amplitude by a factor of only $10\%$ to $20\%$.  

The computed wave functions are reported in the Appendix.  Their
analysis shows the strong contributions of $O^-$ configurations on the
bridging oxygen in the $\bot$ geometries and moderate ones in the $\|$
geometries. These configurations are negligible in the prime geometry,
where their weight is smaller than $1\%$.  This is easily understood
by the quasi-orthogonal character of the angles between the Vanadium
atoms and the two bridging Oxygens, this quasi-orthogonality strongly
hinders the overlap between the V magnetic orbitals and the $p$
orbitals of the bridging oxygens. Finally it results in the
ferromagnetic character of $J^\prime$ and the weakness of the
$t^\prime$ hopping integral.  \\ In the rung or orthogonal geometry,
the contribution of the $O^-$ configurations is particularly large
with weights of $15\%$ in the triplet state, $25\%$ in the singlet
state and weights as large as $49\%$ ($D_{-}$) and $53\%$ ($D_{+}$) in
the two doublet states. This tremendous delocalization of the magnetic
electron on the $p$ orbital of the bridging oxygen is responsible for
the accordingly large hopping integral $t_\bot$, indeed the direct (by
opposition to the through bridge) hopping mechanism contributes only
by a few $meV$ to the total amplitude. Similarly the strong $O^{-}$
contributions in the singlet and the triplet states induce a large
super-exchange mechanism responsible for the strong anti-ferromagnetic
character of $J_\bot$.  Different reasons can be pleaded to explain
these very large delocalization of the magnetic electrons on the
bridging oxygen. The most important is the relative positions of the
pyramids with respect to the bridging O atoms. The vanadium magnetic
$3d$ orbitals are orthogonal to the $V=O$ bonds and the angle of the
$V=O$ bond to the oxygen plane is closed, the corresponding tilting of
the vanadium magnetic orbitals toward the oxygen plane enhances
considerably the overlap between these orbitals and the $p$ orbitals
of the bridging oxygen. Another important factor is the short distance
($1.83\AA$) between the vanadium atoms and the bridging oxygen in this
geometry.  \\ The $\|$ geometry is slightly different since two $p$
orbitals (of different symmetries) of the bridging oxygen are involved
in the through bridge processes. Their consolidate weights in the
doublet states are $20\%$ in $D_{-}$ and $26\%$ in $D_{+}$ and they
result in a moderate enhancement of $t\|$. In the singlet and triplet
states however, the $O^{-}$ configurations do not contribute in a
significant way to the wave functions since their consolidated weight is
smaller than $1\%$.  The result is a very weak super-exchange
mechanism resulting in a slightly anti-ferromagnetic effective exchange
integral ($J_\| \simeq J_\bot/ 60$).

\begin{figure}[h]
\hspace{3cm} \includegraphics{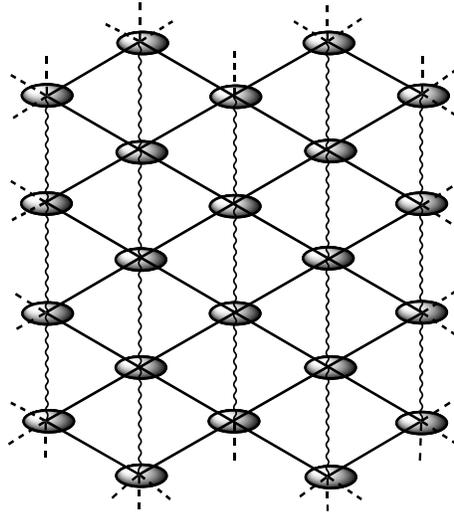} 
\caption{Equivalent magnetic system. The ellipsoids represent the
delocalized magnetic electron on the rungs. The straight lines represent
the ferromagnetic remaining interactions ($J^\prime_{e\!f\!f}$) while the
zig-zag lines represent the anti-ferromagnetic ones ( $J_\|^{e\!f\!f}$).}
\label{Fig4} 
\end{figure}

The dominant interactions are by far those taking place on the ladder
rungs ($t_\bot = -538.2 meV$ and $J_\bot = -293.5 meV$), thus they
should determinate the main representation of the electronic
structure. Two different descriptions can be proposed depending which
of the exchange or the delocalization processes shall dominate. Either
two electrons are paired in a singlet, one rung out of two (favored by
the large value of $J_\bot$), or an unpaired electron is delocalized
on each rung (favored by the large hopping integral $t_\bot$). A
simple energetic calculation shows however that the second solution is
much more favorable ---~with $E_2 / N_{rungs} = t_\bot = -538.2
meV$~--- than the first one ---~$E_1 / N_{rungs} = 1/2 J_\bot =
-147meV$. On the basis of such a representation, the $\alpha^\prime
N\!aV_2O_5$ system can be seen as a two dimensional triangular
Heisenberg system (see fig.~\ref{Fig4}) where the effective magnetic
sites are delocalized on the $V-O-V$ rungs, in agreement with the
representation suggested by Horsch and Mack~\cite{Horsch}. A
wave-functions analysis of the two many-body states involved in the
$t_\bot$ calculations confirm this result since the determinants
involving an unpaired electron on the bridging oxygen are as probable
as the determinants where the magnetic electron is located on the
vanadiums. Indeed their relative weights are $1.00$ in the $D_{-}$ and
$1.21$ in the $D_{+}$.  The exchange integrals between these effective
sites are then anti-ferromagnetic in the $b$ direction, $J_\|^{e\!f\!f} =
J_\|/2$, and ferromagnetic between the effective chains (representing
the ladders), $J^\prime_{e\!f\!f} = J^\prime /4$.


\section{The extended Hubbard model}

\subsection{Coherence of the $t-J$ parameters}

One of the first reflexes in front of quantitative evaluations of
effective integrals in a $t-J$ model is to check the following relation 
\begin{eqnarray} 
\label{rap}
{J_\bot \over J_\|} &=& \left( {t_\bot \over t_\|} \right)^2 
\end{eqnarray}
A quick calculation, using table~\ref{tab:tJres} results, shows that the
above relation is far from being verified in our case. Why?

Equation~\ref{rap} comes from a perturbative evaluation, $J = -4t^2/U_d$,
of the exchange integral from an underlying Hubbard or extended
Hubbard model, supposed to be the exact representation. We saw in the
previous section that the contributions of the $O^{-}$ configurations
are tremendously large in the rung geometry, a Hubbard model based
only on the magnetic $d$ orbitals of the vanadium atoms is therefore
not realistic and any representation pretending to be {\em the reference}
should include the $p$ orbital of the bridging oxygen (at least in
the rung geometry). 

We will use the following notations:  subscripts $p$ and $d$
respectively refer to the $p$ orbital of the bridging oxygen and the
magnetic $d$ orbital of the vanadium.  $\delta=\epsilon_d
-\epsilon_p$ is the orbital energy difference. $U_p$ and $U_d$ are
the on-site repulsion energies. $V_{pd}$ is the vanadium-oxygen
nearest neighbor coulombic repulsion.  The hopping integrals require a little
more care since they can be encountered  in different situations
according to the number and nature of the surrounding electrons: 
\begin{eqnarray*}
\langle \bar{p}| H |\bar{d}\rangle 
&=&\langle \bar{p}|h|\bar{d}\rangle \\ 
\langle d\bar{p}| H |d\bar{d}\rangle &=& 
\langle \bar{p}| h |\bar{d}\rangle + \langle d\bar{p}| V |d\bar{d}\rangle \\
\langle p\bar{p}| H |p\bar{d}\rangle &=& 
\langle \bar{p}|h|\bar{d}\rangle + \langle p\bar{p}| V |p\bar{d}\rangle \\ 
\langle dp\bar{p}| H |dp\bar{d}\rangle &=& 
\langle \bar{p}| h |\bar{d}\rangle + \langle p\bar{p}| V |p\bar{d}\rangle + 
\langle d\bar{p}| V |d\bar{d}\rangle
\end{eqnarray*}
where $h$ refers to the mono-electronic part (kinetic and nuclear
attraction) of the Hamiltonian and $V=1/r_{12}$ is the bi-electronic
repulsion. \\ The $p$ and $d$ orbitals being localized on nearest
neighbor atoms, the $\langle p\bar{p}| V |p\bar{d}\rangle$ and
$\langle d\bar{p}| V |d\bar{d}\rangle$ bi-electronic integrals are a
priori of the same order of magnitude as the mono-electronic part of
the hopping $\langle \bar{p}| h |\bar{d}\rangle$~; thus one should
consider 4 different types of hopping integrals according to the
number and the nature of the spectator electrons on the bond where the
hopping takes place. Let us define
\begin{eqnarray*}
t &=&\langle \bar{p}| H |\bar{d}\rangle  \\
t^d&=&\langle d\bar{p}| H |d\bar{d}\rangle \\
t^p&=&\langle p\bar{p}| H |p\bar{d}\rangle \\
t^{pd}&=&\langle dp\bar{p}| H |dp\bar{d}\rangle 
\end{eqnarray*}

Within such a formalism the leading perturbative term of the
 effective exchange integral between the $i$ and $j$ vanadium
 sites comes at the fourth order of perturbation
\begin{eqnarray} 
\label{jpert}
 J_\bot &=& -4 {\left( t^d t^{pd} \right)^2 \over \left(\Delta_1\right)^2
	\Delta_2} - 8 {\left( t^d t^{pd} \right)^2 \over
	\left(\Delta_1\right)^2 U_d} \\
   &=&  4J_{1,\bot} + 8J_{2,\bot}
\end{eqnarray}
where $\Delta_1 = \delta - U_p + U_d - V_{pd}$ and $\Delta_2 = 2\delta
- U_p + 2U_d - 4V_{pd}$. 

The first term in the expression of $J_\bot$ comes
from the configuration $\left(|d_i \bar{d_i} p \bar{p} \rangle + |p
\bar{p}d_j \bar{d_j} \rangle\right) / \sqrt{2}$ and while second comes
from $|d_i\bar{d_i} d_j \bar{d_j}\rangle$. The second term is usually
negligible in front of the first one, while that one is equal to
$-4t_\bot^2/U_d$. However, due to the strong contributions of the magnetic
electron delocalization on the bridging oxygen, this is not presently
true.  The ratio between $J_{1,\bot}$ and $J_{2,\bot}$ can be
evaluated from the singlet wave functions
coefficients. Indeed the second order perturbative expression of the
singlet wave function is given by
\begin{eqnarray*} 
|\psi_{Sg}\rangle &=& \quad
{|d_i p\bar{p}\bar{d_j}\rangle - |\bar{d_i}p\bar{p}d_j \rangle \over \sqrt{2}} 
\quad + \\ && C_1
{|d_i\bar{d_i}p\bar{d_j}\rangle - |d_i\bar{d_i}\bar{p}d_j\rangle
-|d_i\bar{p}d_jp\bar{d_j}\rangle + |\bar{d_i}pd_jp\bar{d_j}\rangle \over 2} 
\quad + \\ && C_2
{|d_i\bar{d_i}p\bar{p}\rangle + p\bar{p}|d_j\bar{d_j}\rangle \over \sqrt{2}} \quad + \quad 
C_2^\prime
|d_i\bar{d_i}d_j\bar{d_j}\rangle  \\
{\rm where} && \\
C_1 &=& {\sqrt{2} t^{pd} \over \Delta_1} \\
C_2 &=& {2 t^p t^{pd} \over \Delta_1 U_d}  \\
C_2^\prime &=& {2\sqrt{2}t^d t^{pd} \over \Delta_1 \Delta_2} 
\end{eqnarray*}
Thus $(4J_{1,\bot})/(8J_{2,\bot}) = \sqrt{2} C_2 / 2 C_2^\prime \simeq 1.11$
(see Appendix). It comes $4J^1_\bot \simeq -154.3 meV$.

In the ladder direction the $O^{-}$ configurations have negligible
weights in the singlet and triplet states, thus the effective exchange
integral can be expressed as usual~: $J_\| = 2K_\| - 4t_\|^2/ U_d =
2K_\| + 4J_{1,\|}$, where $K_\|$ is the direct exchange integral. The
latter is usually considered as negligible, however the extremely
small value of $J_\|$ imposes to explicitly take these effects into
account in the present case. A direct evaluation yields $K_\|=2.0meV$
and the super-exchange contribution comes to be $4J_{1,\|}=9.1meV$.  One
can now properly verify the coherence of our model hopping and exchange parameters~:
\begin{eqnarray*}
\left({t_\bot \over t_\|}\right)^2 = 18.7 \quad &\simeq&\quad 
{4J_{1,\bot} \over 4J_{1,\|}} = 17.0
\end{eqnarray*}

>From the above analysis it is clear that any method that would not
explicitly take into account the delocalization on the bridging oxygen
would obtain doubtful evaluations of the effective integrals in the
present case. Indeed Smolinski {\em et al}~\cite{Smolinski} found
hopping integrals in agreement with our results using the LDA+U
approach, however their exchange integrals are very different to ours.
This discrepancy can be attributed to the fact that they do not take
into account the delocalization of the magnetic electron on the
bridging oxygen that accounts for half of $J_\bot$ value.

\subsection{The extended Hubbard Hamiltonian}

In addition to the $t-J$ model, it is possible to obtain from our
energy and wave function calculations, the parameters for the extended
Hubbard model described in the previous section~; that is a model
based on the $d$ magnetic orbitals of the vanadium atoms and the $p$
orbitals of the bridging oxygen in the rung geometry. The parameter
extraction will be done using a least square fit method where both the
energy differences and the wave functions coefficients of the extended
Hubbard model are fitted on the computed {\em ab initio} wave
functions and energies, according to the intermediate Hamiltonian
theory~\cite{Hami}.  The different parameters of the Hubbard
Hamiltonian are therefore optimized so that the Hubbard secular
equations of the singlet, triplet and the two doublets are verified
with the computed wave functions (taken as the normalized projection
onto the CAS of the $DDCI2$ wave functions) and the computed energy
differences. The resulting parameters in the rung geometry are
summarized in table~\ref{hubres}.

These values are in reasonable agreement with the vanadium on site repulsion
obtained from the ladder geometry 
$$U_d = 4\left(t_\|\right)^2 / \left( 2K_\| - J_\| \right) = 6.8eV$$

\section{Conclusion}

We computed the effective hopping and exchange integrals of a $t-J$
Hamiltonian for the high temperature phase of $\alpha^\prime N\!a V_2
O_5$. The parameters evaluations were performed using an embedded
cluster approach devised to properly take into account the
electrostatic and exclusion effects of the rest of the crystal as well
as dynamical polarization and correlation effects within the
cluster. The results yield dominant interactions to be the
delocalization of the magnetic electrons on the ladder rungs. This
incredibly large delocalization is mediated by the appropriate $p$
orbital of the bridging oxygen. In fact the magnetic electrons should
not be seen as supported by the $d$ orbitals of the vanadium atoms in
a quarter-filled system, but by orbitals delocalized on the 3 atoms of
the rungs, that is on the $d$ orbitals of the two vanadium atoms and
the $p$ orbital of the bridging oxygen. In such a representation the
system is no longer quarter-filled but half-filled. The remaining
interactions between these delocalized magnetic electrons devise a
two-dimensional triangular Heisenberg system where the effective
exchange in the ladder direction is anti-ferromagnetic while it is
ferromagnetic in the two other directions. 

In addition to the $t-J$ parameters evaluation, a thorough analysis
of the {\em ab-initio} variational wave functions allowed us to evaluate the
relative amplitudes of the underlying three-bands extended Hubbard 
Hamiltonian. The negligible difference obtained between the oxygen
on-site repulsion and the $p-d$ orbital energy difference, $U_p -
\delta < 0.03eV$, yields a quasi-degeneracy between the configurations
where the magnetic electron is located on one of the $d$ orbitals of
the vanadium atoms and the $p$ orbital of the bridging oxygen. It 
 results in the delocalization of the magnetic electrons on the
rungs.

\acknowledgements We thank Jean-Pierre Daudey, Jean-Paul Malrieu,
Fr\'ed\'eric Mila and Timothy Ziman for many fruitful discussions,
T. Chatterji for crystallographic data, Thierry Leininger and Daniel
Maynau for computational help and development. The CASCSF calculations
where done using the MOLCAS package~\cite{MOLCAS}, the DDCI
calculations where done using the CASDI package written by Daniel
Maynau and Nadia Ben Amor~\cite{CASDI}. Part of the calculations were
performed at IDRIS/CNRS under project number 1104.

\section*{Appendix}
The wave functions obtained at the DDCI2 level are developed onto wave functions including up to 592000 slater determinants (according to the state and the
geometry), nevertheless their major contributions are within the CAS
space. The CAS configurations of the computed wave functions are
reported below, after relocalization of the optimized CAS orbitals.
This was done by the mean of a procedure based on Boys'
method~\cite{BOYS}.

In the rung geometry
\begin{eqnarray*} 
|\phi_{Sg} \rangle &=&
0.84 {|d_ip\bar{p}\bar{d_j}\rangle-|\bar{d_i}p\bar{p}d_j\rangle \over 
\sqrt{2}} + \\
&& 0.47{|d_i\bar{d_i}p\bar{d_j}\rangle-|d_i\bar{d_i}\bar{p}d_j\rangle
-|d_i\bar{p}d_j\bar{d_j}\rangle+|\bar{d_i}pd_j\bar{d_j}\rangle \over 2} + \\
&& 0.14{|d_i\bar{d_i}p\bar{p}\rangle+|p\bar{p}d_j\bar{d_j}\rangle \over 
\sqrt{2}}  + 0.09|d_i\bar{d_i}d_j\bar{d_j}\rangle\\ 
&& + {\rm small\; terms\;} \ldots  \\
&&\\
|\phi_{Tp} \rangle &=& 
0.90 {|d_ip\bar{p}\bar{d_j}\rangle + |\bar{d_i}p\bar{p}d_j\rangle \over \sqrt{2}} +\\
&& 0.38{|d_i\bar{d_i}p\bar{d_j}\rangle +|d_i\bar{d_i}\bar{p}d_j\rangle-
|d_i\bar{p}d_j\bar{d_j}\rangle-|\bar{d_i}pd_j\bar{d_j}\rangle \over 2} \\
&& + {\rm small\; terms\;} \ldots  \\
&&\\
|\phi_{D_{+}} \rangle &=&
0.62{|d_ip\bar{p}\rangle +|p\bar{p}d_j\rangle \over \sqrt{2}} + \\
&& 0.69{-2|d_i\bar{p}d_j\rangle +| \bar{d_i}pd_j\rangle+
|d_ip\bar{d_j}\rangle \over \sqrt{6}} + \\ 
&& 0.12{|d_i\bar{d_i}d_j\rangle+| d_id_j\bar{d_j}\rangle \over \sqrt{2}} + \\
&&0.12{|d_i\bar{d_i}p\rangle-|pd_j\bar{d_j}\rangle \over \sqrt{2}}\\ 
&& + {\rm small\; terms\;} \ldots  \\
&&\\
|\phi_{D_{-}}\rangle &=&
0.65{|d_ip\bar{p}\rangle-|p\bar{p}d_j\rangle \over \sqrt{2}} - \\
&& 0.65{|\bar{d_i}pd_j\rangle-| d_ip\bar{d_j}\rangle \over \sqrt{2}} - \\ 
&& 0.06{|d_i\bar{d_i}d_j\rangle-|d_id_j\bar{d_j}\rangle \over \sqrt{2}} + \\
&& 0.11 {|d_i\bar{d_i}p\rangle + |pd_j\bar{d_j}\rangle \over \sqrt{2}} \\
&& + {\rm small\; terms\;} \ldots  
\end{eqnarray*}

In the ladder geometry 
\begin{eqnarray*}
|\phi_{Sg} \rangle &=& 0.92
{|d_ip_x\bar{p_x}p_y\bar{p_y}\bar{d_j}\rangle -
|\bar{d_i}p_x\bar{p_x}p_y\bar{p_y}d_j\rangle \over \sqrt{2}} + {\rm
small\; terms\;} \ldots \\ &&\\ |\phi_{Tp} \rangle &=& 0.92
{|d_ip_x\bar{p_x}p_y\bar{p_y}\bar{d_j}\rangle +
|\bar{d_i}p_x\bar{p_x}p_y\bar{p_y}d_j\rangle \over \sqrt{2}} + {\rm
small\; terms\;} \ldots \\ &&\\ |\phi_{D_{+}} \rangle &=&
-0.78{|d_ip_x\bar{p_x}p_y\bar{p_y}\rangle +
|p_x\bar{p_x}p_y\bar{p_y}d_j\rangle \over \sqrt{2}} \\ && +
0.46{-2|d_i\bar{p_x}p_y\bar{p_y} d_j\rangle +
|\bar{d_i}p_xp_y\bar{p_y}d_j\rangle+
|d_ip_xp_y\bar{p_y}\bar{d_j}\rangle \over \sqrt{6}} \\ && -
0.15{|\bar{d_i}p_x\bar{p_x}p_yd_j\rangle-|d_ip_x\bar{p_x}p_y\bar{d_j}\rangle
\over \sqrt{2}} \\ && +
0.07{|d_i\bar{d_i}p_xp_y\bar{p_y}\rangle-|p_xp_y\bar{p_y}d_j\bar{d_j}\rangle
\over \sqrt{2}} \\ && - 0.04
{|d_i\bar{d_i}p_x\bar{p_x}p_y\rangle-|p_x\bar{p_x}p_yd_j\bar{d_j}\rangle
\over \sqrt{2}} \\ && + {\rm small\; terms\;} \ldots \\ &&\\
|\phi_{D_{-}}\rangle &=&
+0.82{|d_ip_x\bar{p_x}p_y\bar{p_y}\rangle-|p_x\bar{p_x}p_y\bar{p_y}d_j\rangle
\over \sqrt{2}} \\ && -
0.31{|\bar{d_i}p_xp_y\bar{p_y}d_j\rangle-|d_ip_xp_y\bar{p_y}\bar{d_j}\rangle
\over \sqrt{2}} \\ && - 0.28{-2|d_ip_x\bar{p_x}\bar{p_y} d_j\rangle +
|\bar{d_i}p_x\bar{p_x}p_yd_j\rangle+
|d_ip_x\bar{p_x}p_y\bar{d_j}\rangle \over \sqrt{6}} \\ && -0.08
{|d_i\bar{d_i}p_xp_y\bar{p_y}d_j\rangle+|d_ip_xp_y\bar{p_y}d_j\bar{d_j}\rangle
\over \sqrt{2}}  \\ && +0.04 {|d_i\bar{d_i}p_x\bar{p_x}p_y\rangle -
|p_x\bar{p_x}p_yd_j\bar{d_j}\rangle \over \sqrt{2}} \\ && + {\rm small\; terms\;}
\ldots
\end{eqnarray*}

\newpage

\begin{center}
\begin{table}
\caption{Exchange and hopping parameters in $meV$ for a $t-J$ Hamiltonian
computed at different levels of calculation.}
\bigskip
\begin{tabular}{cddd}
Calculation level & CASCI & IDDCI1 & DDCI2 \\
Physics included & Valence shell &
+dynamical  & +dynamical\\
&&polarization& polarization and correlation\\
\hline
$J^\prime$ & +5.48  & +4.14  & +4.87 \\
$t^\prime$& +37.5 & +28.7  & +44.2 \\
\hline
$J_\|$& -1.05 & -4.64 & -5.04 \\
$t_\|$& -115.8 & -176.5  & -124.6 \\
\hline
$J_\bot$& -60.6 & -321.1  & -293.5 \\
$t_\bot$& -420.6 & -542.7 & -538.2 \\
\end{tabular}
\label{tab:tJres}
\end{table} 

\begin{table}
\caption{Parameters in eV of the extended Hubbard Hamiltonian in the
rung geometry.}
\bigskip
\begin{tabular}{ccccc}
\hline $t^p_\bot$&$t^d_\bot$&$t^{pd}_\bot$&$U_d-V_{pd}$&$U_p-\delta$ \\ 
\hline \\
1.1 & 2.0 &1.3&3.6& $<.03$ \\ \hline
\end{tabular}
\label{hubres}
\end{table}
\end{center}

\newpage

\newpage

\newpage

\newpage

\newpage

\end{document}